\documentclass[twocolumn, usenatbib]{aastex631}

\providecommand{\kms}{\,\ensuremath{\rm{km\,s}^{-1}}}
\newcommand{\cmjj}{\mbox{${\rm cm^{-2}}$}}
\newcommand{\cmjjj}{\mbox{${\rm cm^{-3}}$}}

\newcommand{\msol}{\mbox{${\rm M}_\odot$}}
\newcommand{\mstar}{\mbox{${\rm M}_{\rm star}$}}

\accepted{for publication in ApJL on September 10, 2023}

\shorttitle{}
\shortauthors{Chen et al.}

\begin{document}

\title{The Cosmic Ultraviolet Baryon Survey: Empirical Characterization of Turbulence in the Cool Circumgalactic Medium}

\correspondingauthor{Hsiao-Wen Chen}
\email{hchen@astro.uchicago.edu}

\author[0000-0001-8813-4182]{Hsiao-Wen Chen}
\affil{Department of Astronomy \& Astrophysics, The University of Chicago, 5640 S Ellis Ave., Chicago, IL 60637, USA}

\author[0000-0002-2941-646X]{Zhijie Qu}
\affiliation{Department of Astronomy \& Astrophysics, The University of Chicago, 5640 S Ellis Ave., Chicago, IL 60637, USA}

\author{Michael Rauch}
\affiliation{The Observatories of the Carnegie Institution for Science, 813 Santa Barbara Street, Pasadena, CA 91101, USA}

\author{Mandy C.\ Chen}
\affil{Department of Astronomy \& Astrophysics, The University of Chicago, 5640 S Ellis Ave., Chicago, IL 60637, USA}

\author[0000-0001-7869-2551]{Fakhri S.\ Zahedy}
\affiliation{The Observatories of the Carnegie Institution for Science, 813 Santa Barbara Street, Pasadena, CA 91101, USA}

\author[0000-0002-2662-9363]{Sean D.\ Johnson} 
\affiliation{Department of Astronomy, University of Michigan, Ann Arbor, MI 48109, USA}

\author[0000-0002-0668-5560]{Joop Schaye}
\affiliation{Leiden Observatory, Leiden University, PO Box 9513, NL-2300 RA Leiden, the Netherlands}

\author[0000-0002-8459-5413]{Gwen C.\ Rudie}
\affiliation{The Observatories of the Carnegie Institution for Science, 813 Santa Barbara Street, Pasadena, CA 91101, USA}

\author[0000-0003-3244-0409]{Erin Boettcher}
\affiliation{Department of Astronomy, University of Maryland, College Park, MD 20742, USA}
\affiliation{X-ray Astrophysics Laboratory, NASA/GSFC, Greenbelt, MD 20771, USA}
\affiliation{Center for Research and Exploration in Space Science and Technology, NASA/GSFC, Greenbelt, MD 20771, USA}

\author[0000-0001-5804-1428]{Sebastiano Cantalupo}
\affiliation{Department of Physics, University of Milan Bicocca, Piazza della Scienza 3, I-20126 Milano, Italy}

\author[0000-0002-4900-6628]{Claude-Andr\'e Faucher-Gigu\`ere}
\affiliation{Department of Physics \& Astronomy, Center for Interdisciplinary Exploration and Research in Astrophysics (CIERA), Northwestern University, 1800 Sherman Avenue, Evanston, IL 60201, USA}

\author[0000-0002-5612-3427]{Jenny E.\ Greene}
\affiliation{Department of Astrophysics, Princeton University, Princeton, NJ 08544, USA}

\author{Sebastian Lopez}
\affiliation{Departamento de Astronom\'{i}a, Universidad de Chile, Casilla 36-D, Santiago, Chile}

\author{Robert A.\ Simcoe}
\affiliation{MIT-Kavli Institute for Astrophysics and Space Research, 77 Massachusetts Ave., Cambridge, MA 02139, USA}







\begin{abstract}

This paper reports the first measurement of the relationship between turbulent velocity and cloud size 
in the diffuse circumgalactic medium (CGM) in typical galaxy halos at redshift $z\approx 0.4$\,-\,1. 
Through spectrally-resolved absorption profiles of a suite of ionic transitions paired with careful ionization analyses of individual components, 
cool clumps of size as small as $l_{\rm cl}\sim 1$ pc and density lower than $n_{\rm H}=10^{-3}\,\cmjjj$ are identified in galaxy halos.  In addition, comparing the line widths between different elements for kinematically matched components provides robust empirical constraints on the thermal temperature $T$ and the non-thermal motions $b_{\rm NT}$, independent of the ionization models.  
On average, $b_{\rm NT}$ is found to increase with $l_{\rm cl}$ following $b_{\rm NT}\propto l_{\rm cl}^{0.3}$ over three decades in spatial scale from $l_{\rm cl}\approx 1$ pc to $l_{\rm cl}\approx 1$ kpc. 
 Attributing the observed $b_{\rm NT}$ to turbulent motions internal to the clumps, the best-fit $b_{\rm NT}$-$l_{\rm cl}$ relation shows that the turbulence is consistent with Kolmogorov at $<1$ kpc with a roughly constant energy transfer rate per unit mass of $\epsilon\approx 0.003\,{\rm cm}^{2}\,{\rm s}^{-3}$ and a dissipation time scale of $\lesssim\,100$ Myr.  No significant difference is 
 found 
 between massive quiescent and star-forming halos in the sample on scales less than 1 kpc.  While the inferred $\epsilon$ is comparable to what is found in \ion{C}{4} absorbers at high redshift, it is considerably smaller than observed in star-forming gas or in extended line-emitting nebulae around distant quasars. A brief discussion of possible sources to drive the observed turbulence in the cool CGM is presented.


\end{abstract}




\section{Introduction}

\label{sec:intro}


The circumgalactic medium (CGM) at the interface between star-forming regions and intergalactic space contains the critical record of gas circulation in and out of galaxies. Tremendous progress has been made over the last decade in establishing a physical understanding of this diffuse gas.  The CGM has been shown to be kinematically complex \citep[e.g.,][]{Rudie:2019,Zahedy:2019, Qu:2023}, clumpy and multiphase, spanning a broad range in density, temperature, and metallicity \citep[see e.g.,][for reviews]{Chen:2017,Donahue:2022}.  The dynamic state of the gas is dictated by different feeding and feedback processes and is intimately connected to how galaxies grow and evolve \citep[see e.g.,][for a review]{Faucher:2023}.  In particular, the cool CGM with a typical gas density of $n_{\rm H}\!\approx\!0.01\,\cmjjj$ and temperature of $T\!\approx\!10^4$ K \citep[e.g.,][]{Qu:2022}, a nominal source of fuel for sustaining star formation, has an expected effective mean free path of $\sim\!10^{14}$ cm and kinematic viscosity of $\approx\!10^{20}\,{\rm cm}^2\,{\rm s}^{-1}$ \citep[e.g.][]{Spitzer:1962, Sarazin:1986}.  For a typical cool cloud of size $100$ pc \citep[e.g.][]{Zahedy:2021}, moving at a speed of $100$ \kms\ in galaxy halos \citep[e.g.,][]{Huang:2021}, the Reynolds number associated with the gas flow is large, $Re\sim 3\times 10^7$ \citep[cf.][for cold/warm neutral medium observed in Milky Way high-velocity clouds]{Marchal:2021}, showing that the cool CGM clouds should be turbulent \citep[see e.g.,][for a recent discussion on the turbulent nature of diffuse ionized gas]{Burkhart:2021}.

Turbulent energy can provide additional heating beyond feedback from stars and active galactic nuclei (AGN) to offset cooling in the hot halo through non-linear interactions between large and small eddies \citep[e.g.,][]{McNamara:2007,Zhuravleva:2014}.  In addition, turbulence produces density fluctuations, triggering multiphase condensation in the hot halo \citep[e.g.,][]{Voit:2017,Gaspari:2018}.  Furthermore, turbulent mixing also provides an efficient transport mechanism for metals from star-forming regions to the CGM and possibly beyond \citep[e.g.,][]{Pan:2010}.  

Despite the vital scientific implications, robust empirical constraints of turbulence on large scales in diffuse halo gas remain scarce \citep[see e.g.,][]{Rauch:2001,Rudie:2019,Li:2020,Chen:2023}.  Here we report the first measurement of the turbulent velocity-size relation in the diffuse CGM around typical $L_*$ and sub-$L_*$ galaxy halos at intermediate redshifts, $z\approx 0.4-1$, based on observations that utilize high-resolution absorption spectroscopy.
We show that the observed non-thermal velocity width of individually-resolved components, $b_{\rm NT}$, scales with the size of the absorbing clumps, $l_{\rm cl}$, according to $b_{\rm NT}\propto\,l_{\rm cl}^{0.3}$, consistent with the expectation for subsonic turbulence from the Kolmogorov theory \citep[e.g.,][]{Kolmogorov1941,Frisch:1996}.

This paper is organized as follows. Section~\ref{sec:methodology} describes the method adopted for characterizing the turbulence in the diffuse CGM.  Section~\ref{sec:data} summarizes the data collected for component line width and size measurements.   In Section~\ref{sec:result}, we present an empirical relation between the turbulent velocity and size of individual, spectrally-resolved gaseous clumps.  In Section~\ref{sec:discussion}, we discuss the implications of our findings.  

\section{Method} \label{sec:methodology}

To characterize turbulence, a classical approach 
is to measure the velocity structure functions (VSFs) that characterize the scale-dependent variance of the velocity field \citep[e.g.,][]{Frisch:1996,Boldyrev:2002}.  
For a homogeneous, isotropic, and incompressible fluid, the energy transfer between different scales is conserved.  One can infer
based on a dimensional analysis that a constant energy transfer rate would lead to a constant $\Delta\,v^2(l)\times\Delta\,v(l)/l$. 
The observed velocity variance between different locations depends on the distance separation $l$ according to $\langle\,|\Delta\,v(l)|^2\,\rangle\propto l^{2/3}$ and the energy transfer rate per unit mass $\epsilon$ is related to the velocity dispersion according to $\epsilon=(5/4)\langle\,|\Delta\,v(l)|^3\,\rangle/l$ \citep[e.g.,][]{Kolmogorov1941}. 
For subsonic turbulence, kinetic energy injected on large scales is expected to propagate to small scales at a constant rate ($\epsilon$), eventually dissipating at the smallest scale when viscosity transforms the kinetic energy into heat. 
Observations of scale-dependent gas velocity dispersion, therefore, place direct constraints on the turbulent energy cascade in the diffuse CGM.

Spatially-resolved velocity maps of diffuse gas can be established using absorption spectroscopy of multiply-lensed QSOs or line-emitting nebulae observed using integral-field spectroscopy (IFS).  By targeting multiply-lensed QSOs, \cite{Rauch:2001} presented the first VSF measurements of the intergalactic medium at $z\!\approx\!3$ on spatial scales between 30 pc and 30 kpc.  However, the measurements are limited by a still small sample of multiply-lensed QSOs \citep[see also][]{Chen:2014}.  IFS observations of nearby massive galaxy clusters have revealed spatially extended line-emitting gaseous streams in the intercluster medium enabling VSF measurements on scales from 30 pc to 10 kpc \citep[][]{Li:2020}.  Similarly, VSF measurements have also been reported for QSO host halos on scales from $\lesssim\!10$ kpc to $\approx\! 60$ kpc based on line-emitting nebulae discovered around luminous QSOs \citep[][]{Chen:2023}.  However, detections of line-emitting nebulae around typical galaxies or galaxy groups remain extremely rare beyond the nearby universe \citep[e.g.,][]{Chen:2019a,Dutta:2023} and even for extreme systems like QSO nebulae and intracluster medium the ground-based seeing imposes a limit on the smallest scale one can probe with IFS observations.  Extending VSF studies to the CGM around galaxies representative of the field population on scales below 1 kpc requires a different approach.

High-resolution absorption spectra of distant QSOs provide unsurpassed sensitivity for probing the physical and thermodynamic conditions of the diffuse multiphase CGM.  Specifically, gas with different densities and elemental abundances being photo-ionized by an ambient radiation field is expected to exhibit different relative ion abundances \citep[e.g.,][]{Ferland:2017}.  Comparing relative ionic column density ratios of kinematically matched components, therefore, constrains $n_{\rm H}$ and the ionization fraction of different species including hydrogen.  Combining the observed neutral hydrogen column density $N_{\rm H{\small I}}$ and the anticipated neutral fraction of hydrogen, $f_{{\rm HI}}$, yields the 
size of the absorbing clump along the line of sight, $l_{\rm cl}=N_{\rm HI}/(f_{\rm HI}\,n_{\rm H})$.  
Through careful ionization analyses of resolved components, 
it is possible to identify individual clumps of size as small as $l_{\rm cl}\!\sim\!1$ pc and density lower than $n_{\rm H}\!=\!10^{-3}\,\cmjjj$ in galaxy halos \citep[e.g.,][]{Schaye:2007,Zahedy:2019}.  

\begin{figure*}
  \begin{center}
\includegraphics[width=\textwidth]{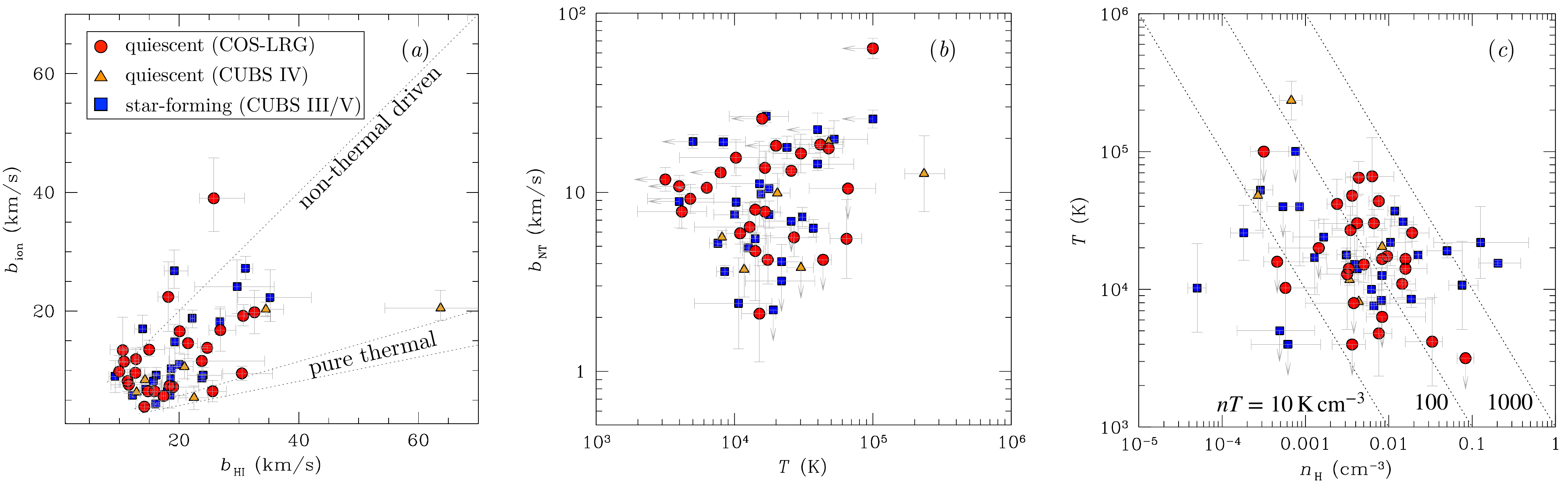}
\end{center}
\caption{Summary of the physical properties of individual components collected from the COS-LRG \citep[e.g.,][solid red circles]{Zahedy:2019} and CUBS \citep[e.g.,][solid orange triangles and solid blue squares]{Zahedy:2021,Cooper:2021,Qu:2022} programs. Panel ({\it a}) displays the measured Doppler parameters ($b_{\rm ion}$) for low-ionization metal species along the y-axis and those of \ion{H}{1} ($b_{\rm HI}$) along the x-axis.  Error bars represent the 68\% confidence interval. The low-ionization lines included in the analysis are primarily the \ion{Mg}{2} doublet recorded in ground-based high-resolution optical echelle spectra, supplemented with far-ultraviolet lines, such as \ion{C}{2} and \ion{O}{2}, from {\it HST} COS observations.  The dotted lines mark the two extreme scenarios of pure non-thermal broadening (top) and pure thermal broadening (bottom for carbon and magnesium).  
Panel ({\it b}) displays the derived constraints on the non-thermal motion $b_{\rm NT}$ and thermal temperature $T$ based on comparisons between the line widths of hydrogen and the ions presented in panel ({\it a}).  Upper limits for $T$ and $b_{\rm NT}$ are shown as left-pointing and downward-pointing arrows, respectively.  Panel ({\it c}) displays the line-width inferred thermal temperature $T$ versus the gas density inferred from photo-ionization models \citep[e.g.,][]{Zahedy:2019,Zahedy:2021,Cooper:2021,Qu:2022}.  Dotted lines mark the expectations from constant thermal pressure models of $nT=10, 100, 1000\,{\rm K\,cm^{-3}}$, suggesting a scatter of more than two decades in the internal gas pressure among individual clumps \citep[see also][]{Qu:2022}.}
\label{fig:summary}
\end{figure*}

In parallel, the observed Doppler line width ($b$) of a resolved absorption component is related to the line-of-sight velocity dispersion $\sigma_v^{\rm los}$ within the absorbing clump according to $b=\sqrt{2}\,\sigma_v^{\rm los}$.  It is a combined measure of the underlying thermal ($b_T$) and non-thermal ($b_{\rm NT}$) motions \citep[e.g.,][]{Rauch:1996}, $b^2=b^2_T+b^2_{\rm NT}$.  Because $b_T=\sqrt{2k_BT/m_I}$, it depends on both gas temperature $T$ and ion mass $m_I$, and more massive particles have smaller $b_T$.  At the same time, $b_{\rm NT}$ is shared among all elements, irrespective of the particle mass.  Comparing the line widths between elements of sufficiently different mass for kinematically matched components, therefore, provides robust empirical constraints to decouple $T$ and $b_{\rm NT}$.  This exercise is entirely independent of the ionization models.  The agreement between the temperature determined from the line widths of different elements and the temperature inferred from the best-fit photo-ionization model \citep[e.g.,][]{Qu:2022} also provides additional support for the robustness of the photoionization modeling, as well as validation for the assumption of ionization and thermal equilibrium. 

Combining $l_{\rm cl}$ determined based on the best-fit ionization model for individual components and $b_{\rm NT}$ 
from line-width analyses provides a unique opportunity to investigate the internal turbulence of resolved cool clumps on sub-kpc scales and how the turbulent velocity differs between clumps of different sizes in the CGM.

\section{Empirical data} \label{sec:data}

To investigate whether a correlation is present between the observed internal turbulence of resolved cool CGM clumps and the inferred clump size, we have assembled a sample of spectrally-resolved CGM absorbing components with available constraints on the thermal and ionization properties.  In particular, the data are collected from the COS-LRG program that characterizes the CGM absorption properties of luminous red galaxies (LRGs) at $z\!=\!0.2$-0.55 \citep[][]{Chen:2018}, as well as absorption constraints for the CGM around field galaxies or galaxy groups at $z\approx 0.4$-1 identified in the Cosmic Ultraviolet Baryon Survey \citep[CUBS;][]{Chen:2020}.  In particular, the LRGs are characterized by luminosities of $\gtrsim\!3\,L_*$ and stellar masses of $\mstar\gtrsim 10^{11}\,\msol$ with little ongoing or recent star formation \citep[e.g.,][]{Roseboom:2006}, while the CUBS sample has the closest galaxies spanning a range in mass from $\mstar\approx 10^8\,\msol$ to $\mstar\approx\,10^{10}\,\msol$ \citep[e.g.,][]{Chen:2020,Qu:2023} and star formation rate averaged over the past 300 Myr from ${\rm SFR_{300}}\approx 1.3\,\msol\,{\rm yr}^{-1}$ to ${\rm SFR_{300}}\lesssim 30\,\msol\,{\rm yr}^{-1}$ \citep[e.g.][]{Qu:2023}.

Both surveys employ high-resolution absorption spectra obtained on the ground and in space for resolving the multi-component line profiles of different ionic transitions.  Specifically, optical echelle spectra obtained using either MIKE on the Magellan Telescopes \citep[][]{Bernstein:2003} or HIRES on Keck \citep[][]{Vogt:1994} provide a full-width-at-half-maximum (FWHM) spectral resolving power of ${\rm FWHM}\approx 5$-10 \kms\ for low-ionization lines including \ion{Mg}{2} and \ion{Fe}{2} \citep[e.g.][]{Zahedy:2019}, while far-ultraviolet (FUV) spectra obtained using the Cosmic Origins Spectrograph \citep[COS;][]{Green:2012} on the {\it Hubble Space Telescope} ({\it HST}) provide a resolution of ${\rm FWHM}\approx 18$ \kms\ for resolving transitions produced by \ion{H}{1} and a suite of FUV lines covering \ion{C}{2}-{\small IV}, \ion{N}{2}-{\small IV}, \ion{O}{2}-{\small VI}, and \ion{Si}{2}-{\small IV} \citep[e.g.][]{Zahedy:2021,Cooper:2021,Qu:2022}.  The combined spectral resolving power and the dynamic range in particle mass enable an accurate decomposition between thermal and non-thermal contributions to the observed component line widths.  The broad coverage of different ions enables robust constraints for the ionization conditions.

Figure \ref{fig:summary} summarizes the properties of these measurements.
The observed $b$ values of different metal ions versus \ion{H}{1} are presented in Figure \ref{fig:summary}{\it a} with data collected from four studies: COS-LRG \citep[][]{Zahedy:2019}, galaxies/galaxy groups associated with four Lyman limit systems of $\log\,N_{\rm HI}/\cmjj>17$ \citep[CUBS\,III;][]{Zahedy:2021}, galaxy groups associated with two partial LLSs of $\log\,N_{\rm HI}/\cmjj\approx 16$-16.5 \citep[CUBS\,IV;][]{Cooper:2021}, and galaxies/galaxy groups with multiphase ionic transitions detected in star-forming galaxies at $z\approx 1$ \citep[CUBS\,V;][]{Qu:2022}\footnote{This line-width analysis was performed using \ion{H}{1} and a subset of metal ions.  In particular, when \ion{Mg}{2} is detected, it is given preference because of the large mass differential between H and Mg and because of the high spectral resolving power associated with \ion{Mg}{2} observations.  When \ion{Mg}{2} is not available, UV transitions detected in COS spectra are used, including \ion{C}{2} and \ion{O}{2} along with \ion{H}{1}.  As shown in \cite{Qu:2022}, repeating the $b_{\rm NT}$ and $T$ inference for \ion{Mg}{2}-bearing components based on UV transitions observed in the COS spectra alone leads to a slight increase in the estimated $b_{\rm NT}$ by $\approx 2$ \kms\ and a corresponding decrease in $T$.  However, the results are consistent within the uncertainties.}.  
It is clear that the observed $b$ values spread across the parameter space bordered by the pure-thermal and pure non-thermal regimes.  

The derived $T$ and $b_{\rm NT}$ of these components are presented in Figure \ref{fig:summary}{\it b}.  In cases where the observed $b$ values are consistent for elements of very different mass, only an upper limit can be placed on $T$, which are shown as left-pointing arrows.  Similarly, when the observed differences in $b$ can be primarily attributed to thermal broadening, only an upper limit can be placed on $b_{\rm NT}$, and these are shown as downward pointing arrows.
As expected, $b_{\rm NT}$ and $T$ are well-decoupled and no correlation is seen between these two quantities.  A generalized Kendall's test, incorporating both measurements and upper limits, returns a correlation coefficient of $r<0.1$, consistent with the expectation that $b_{\rm NT}$ and $T$ are not correlated.  

Finally, Figure \ref{fig:summary}{\it c} summarizes the derived thermal temperature $T$ versus clump density $n_{\rm H}$ inferred from best-fit ionization models.  The dotted lines indicate constant thermal pressure, highlighting a large spread in the internal thermal pressure of these resolved clumps \citep[e.g.,][]{Qu:2022}.

\begin{figure}
  \begin{center}
\includegraphics[width=0.475\textwidth]{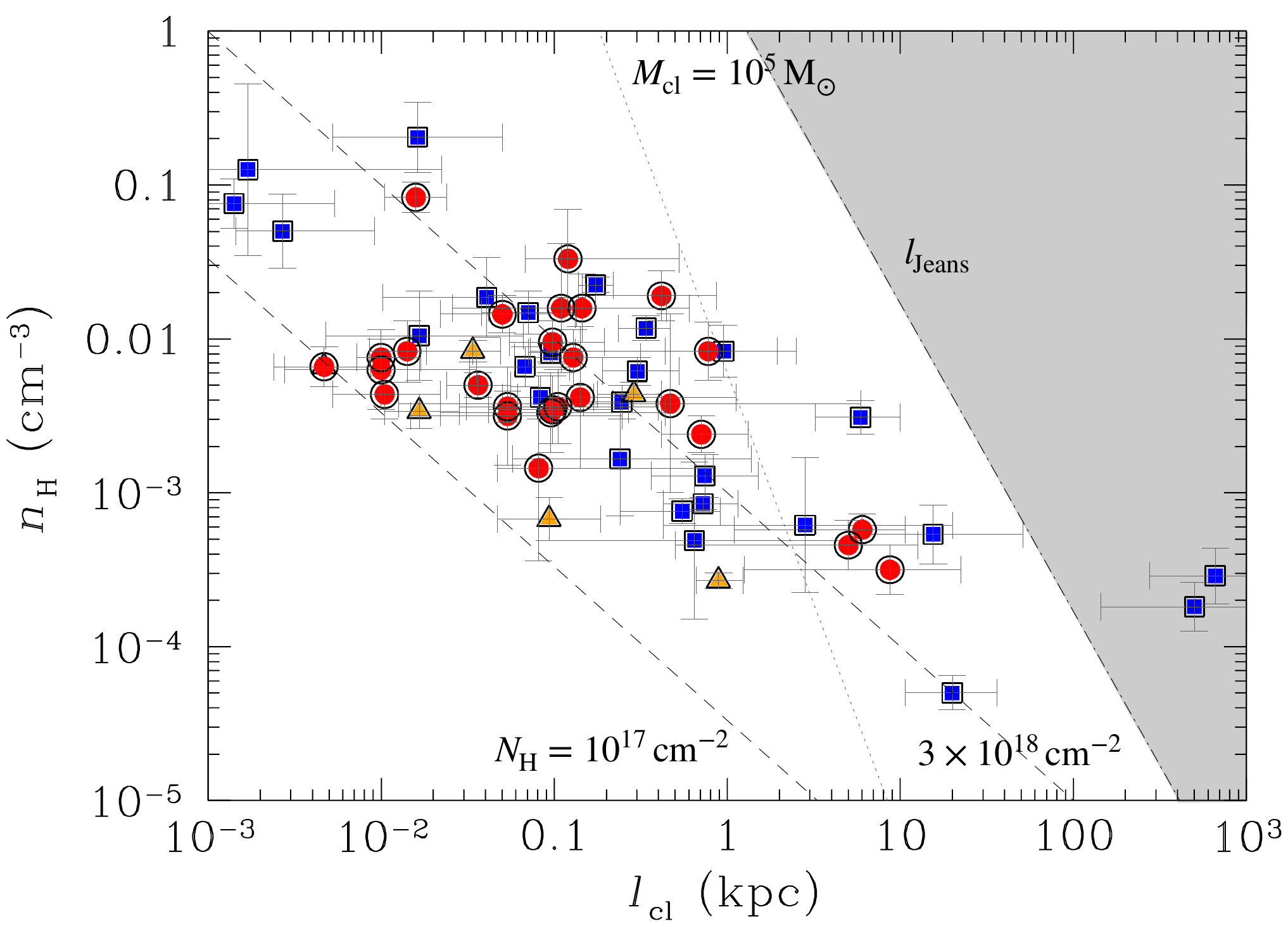}
\end{center}
\caption{Distribution of the inferred gas density $n_{\rm H}$ and clump size $l_{\rm cl}$ for individual components included in this study.  Different symbols indicate the sources of measurements described in Figure \ref{fig:summary} and in the main text.  Long-dashed lines mark the expected $n_{\rm H}$-$l_{\rm cl}$ relation for a fixed $N_{\rm H}$, while the dotted line marks the relation for a constant clump mass of $10^5\,{\rm M}_\odot$.  The shaded area marks the parameter space where the size of the clumps exceeds the Jeans length (the dash-dotted line) and the clumps are expected to be unstable against gravitational collapse.  Note that while survey sensitivities may prevent detections of small, low-density clumps (but see the discussion in \S\ \ref{sec:bias}), the absence of large, high-density clumps is not a result of selection bias.  
}
\label{fig:nsize}
\end{figure}

We further examine the inferred clump size $l_{\rm cl}$ and its correlation with $n_{\rm H}$ from the best-fit ionization models of individual components in Figure \ref{fig:nsize}.    Recall that $n_{\rm H}$ is determined based on column density measurements of a suite of ions (not including \ion{H}{1}) and a fiducial background ionizing radiation field, while $l_{\rm cl}$ is determined according to $l_{\rm cl}=N_{\rm HI}/(f_{\rm HI}\,n_{\rm H})$. 
For each resolved absorbing clump, $l_{\rm cl}$ is, therefore, driven by the observed $N_{\rm HI}$ for a given best-fit photo-ionization model.  

Under a photo-ionization framework, Figure \ref{fig:nsize} shows that $n_{\rm H}$ is inversely correlated with $l_{\rm cl}$ and higher-density gas is confined in smaller clumps.  If the clumps formed at a constant mass, then $n_{\rm H}\!\propto\!l_{\rm cl}^{-3}$ would be expected (dotted line in the figure).  The shallower slope in Figure \ref{fig:nsize} shows that these cool clumps span a broad range in mass.  For comparison, the expectations for constant total hydrogen column density $N_{\rm H}$ at $N_{\rm H}\!=\!10^{17}$ and $3\times 10^{18}\,\cmjj$ are shown in long dashed lines, while the Jeans length ($l_{\rm Jeans}$) inferred for cool gas of $T\approx 10^4 \rm \,K$ and density $n_{\rm H}$ is shown as the dash-dotted line.  It is clear that two unphysically large clumps with $l_{\rm cl}$ exceeding 100 kpc would be unstable against gravitational collapse, suggesting the possible presence of a local ionizing source\footnote{Both clumps have relatively large $N_{\rm HI}$ with $\log\,N_{\rm HI}/\cmjj\approx16$-16.4.  The unphysically large $l_{\rm cl}$ is therefore driven by the low $n_{\rm H}$ inferred based on a fiducial background radiation field.  For a fixed ionization parameter, as constrained by the observed relative ion abundances, a higher ionizing radiation intensity would lead to a higher inferred gas density, reducing $l_{\rm cl}$. The tension between the inferred $l_{\rm cl}$ and $l_{\rm Jeans}$ can, therefore, be alleviated by considering local fluctuations in the ionizing radiation field \citep[e.g.,][]{Zahedy:2021,Qu:2023}.}.  At the same time, the rest of the clumps are all smaller than the Jeans length, implying that either they are pressure confined or short-lived.  In addition, while $N_{\rm H}$ of resolved cool clumps spans a range of roughly two orders of magnitude, no points are found below $N_{\rm H}\!=\!10^{17}\,\cmjj$ \citep[see also][]{Zahedy:2019}.  
We discuss in \S\,\ref{sec:bias} below the nature of this $N_{\rm H}$ threshold, and argue that this is a physical limit as a result of clump formation and survival, rather than selection bias.

\section{Results} \label{sec:result}

\begin{figure}
  \begin{center}
\includegraphics[width=0.475\textwidth]{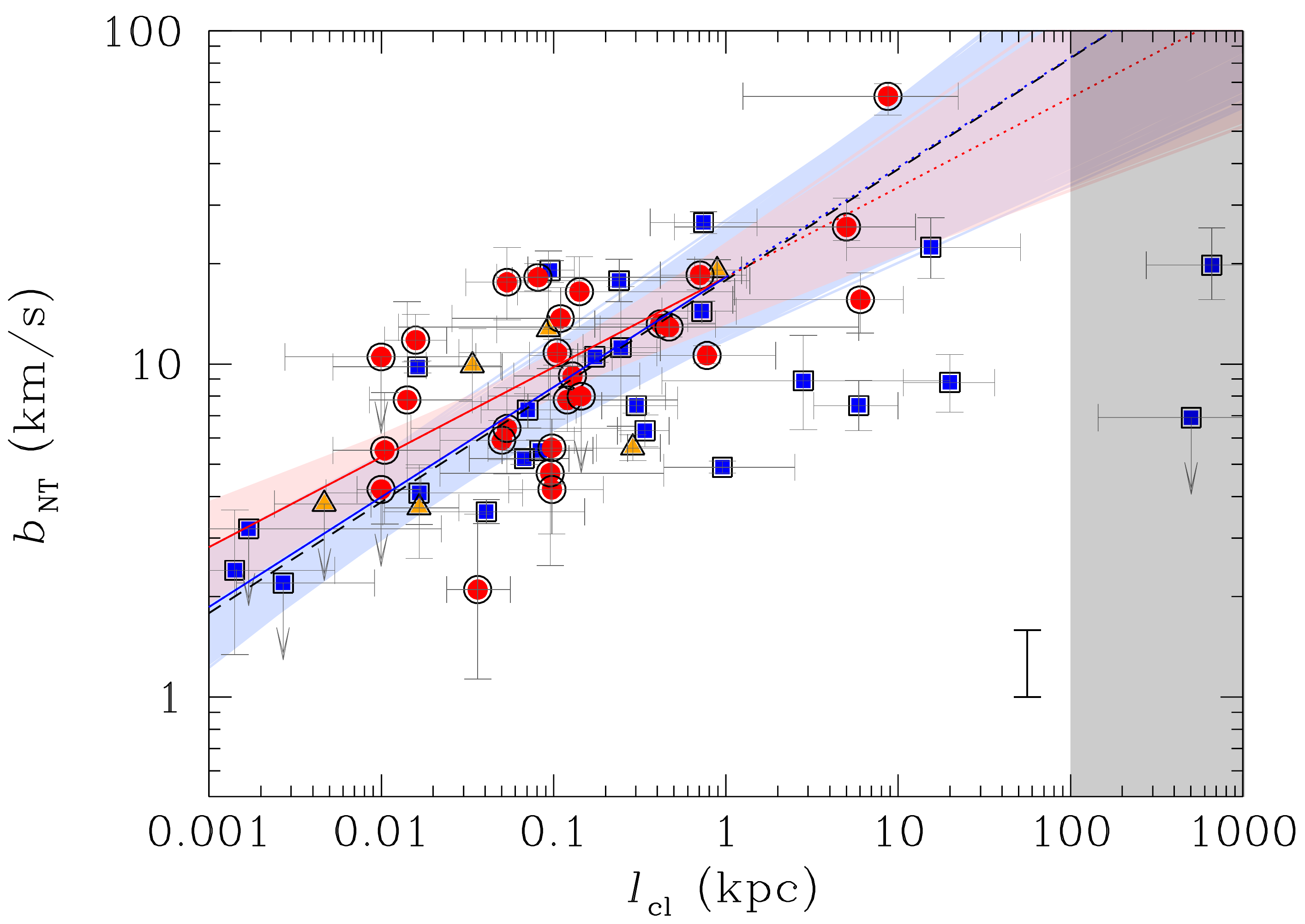}
\end{center}
\caption{Non-thermal line width $b_{\rm NT}$ determined from the observed absorption profiles of individual clumps versus the clump size $l_{\rm cl}$ determined from best-fit photo-ionization models. Different symbols indicate the sources of measurements described in Figure \ref{fig:summary} and in the main text.  Following Figure \ref{fig:summary}, upper limits on $b_{\rm NT}$ (downward arrows) are shown for components exhibiting line widths largely consistent with expectations from the mass-dependent thermal term.  The expectation of $b_{\rm NT}\propto l_{\rm cl}^{1/3}$ from the Kolmogorov theory for subsonic turbulence is shown as the dashed line.  The correlation between $b_{\rm NT}$ and $l_{\rm cl}$ appears to flatten beyond $l_{\rm cl}\approx 1$ kpc in star-forming halos.  The best-fit lines for LRG and star-forming halos at $l_{\rm cl}<1$ kpc are shown in solid red and blue lines, respectively.  The extrapolation to larger spatial scales are shown in dotted line.  Uncertainties associated with the best-fit models are displayed in the corresponding color-shaded area, while the estimated intrinsic scatter $\sigma_{\rm int}$ is indicated by the bar in the lower-right corner.  See text in \S\,\ref{sec:result} for details. 
}
\label{fig:vsf}
\end{figure}

With the thermal and non-thermal velocity widths differentiated by comparing the line widths of different elements
and the sizes of individual clumps determined independently from the observed 
relative column density ratios (\S\,\ref{sec:methodology}), 
we now proceed with investigating how $b_{\rm NT}$ and  $l_{\rm cl}$ are related. In Figure \ref{fig:vsf}, we show the measurements for individual clumps identified in quiescent (red circles and orange triangles) and star-forming (blue squares) halos.  It is clear that larger clumps exhibit on average higher non-thermal motions.  In addition, cool CGM clumps near massive quiescent and star-forming galaxies occupy a similar $b_{\rm NT}$-$l_{\rm cl}$ parameter space.  However, the slope in star-forming halos appears to flatten beyond $l_{\rm cl}=1$ kpc with the observed $b_{\rm NT}$, remaining at $b_{\rm NT}\lesssim 30$ \kms.  

To quantify the correlation between $b_{\rm NT}$ and $l_{\rm cl}$, we adopt a simple power-law model 
\begin{equation}
    \frac{b_{\rm NT}}{b_{\rm NT}^0}=\left(\frac{l_{\rm cl}}{\rm 1\,kpc}\right)^\gamma.
\end{equation} 
We then obtain a best-fit $\gamma$, $b_{\rm NT}^0$, and the associated uncertainties based on a likelihood analysis that incorporates upper limits in the measurements.  We also include an intrinsic scatter $\sigma_{\rm int}$ in the likelihood analysis to account for possible systematic differences between different halos, including possible fluctuations in the ionizing radiation field and in halo density profiles.  We perform the likelihood analysis for clumps identified in quiescent halos and in star-forming ones both separately and as a whole.  Given the apparent flattening trend seen in star-forming halos, we also perform the likelihood analysis using only data points with $l_{\rm cl}<1$ kpc as well as considering all clumps except for the two unphysically large clumps with $l_{\rm cl}>100$ kpc (see \S\ \ref{sec:data}).  The results are summarized in Table \ref{tab:my_label}.  To facilitate subsequent discussions, we have converted the best-fit and associated uncertainties of $b_{\rm NT}^0$ to the corresponding energy transfer rate per unit mass at 1 kpc according to $\epsilon\approx \sigma_v^3/l_{\rm cl}=(\sqrt{3}\,\sigma_v^{\rm los})^3/l_{\rm cl}=(\sqrt{3/2}\,b_{\rm NT})^3/l_{\rm cl}$. This leads to $\epsilon_0=5.95\times 10^{-7}\,(b_{\rm NT}^0/\kms)^3\,{\rm cm}^2\,{\rm s}^{-3}$ at 1 kpc.
The best-fit models obtained separately for quiescent and star-forming halos for clumps with $l_{\rm cl}<1$ kpc are presented in Figure \ref{fig:vsf}, along with the corresponding uncertainties and the estimated $\sigma_{\rm int}$.  
 

\section{Discussion} \label{sec:discussion}

Our analysis shows that the internal dynamics observed in the cool CGM show little distinction between massive quiescent and star-forming halos in available data.  
Recall that the Doppler parameter $b$ is related to the underlying velocity dispersion according to $b\!=\!\sqrt{2}\,\sigma_v^{\rm los}$ and that $b_{\rm NT}$ quantifies non-thermal motions internal to each resolved clump.  In principle, both large-scale bulk flows and turbulent motions can contribute to the observed non-thermal line width.  For these individually resolved absorbing clumps of typical size $l_{\rm cl}\approx 100$ pc, we consider bulk flows an unlikely scenario for the observed $b_{\rm NT}$ because it would imply a large velocity shear over a small volume, exceeding 100 \kms\,kpc$^{-1}$ \citep[cf.][]{Chen:2014}.  Attributing the observed $b_{\rm NT}$ to turbulent motions internal to the clumps, we have an empirical measure of the scale-dependent turbulent velocity field, similar to the first-order VSF \citep[cf.,][]{Li:2020}, for the cool CGM over three decades in spatial scale from $l_{\rm cl}\approx 1$ pc to $l_{\rm cl}\approx 1$ kpc.  

\begin{table}[]
    \centering
    \caption{Summary of best-fit 1st-order VSF slope $\gamma$ and energy transfer rate $\epsilon$ at 1 kpc in Figure \ref{fig:vsf}.}
    \label{tab:my_label}
    \begin{tabular}{lcccc}
    \hline\hline
    & & $\epsilon_0$ & $\sigma_{\rm int}$ & \\
    Sample  & $\gamma$ & ($10^{-3}\,{\rm cm}^2\,{\rm s}^{-3}$) & dex \\
    \hline
    \multicolumn{5}{c}{At $l_{\rm cl}<30$ kpc}\\
    \hline 
    All & $0.25\pm 0.03$ & $2.1_{-0.5}^{+0.8}$ & $0.22\pm0.03$\\
    Star-forming & $0.23_{-0.04}^{+0.05}$ & $1.2_{-0.4}^{+0.7}$ & $0.25_{-0.04}^{+0.05}$\\
    Passive & $0.28\pm0.04$ & $3.9_{-1.3}^{+2.2}$ & $0.17\pm0.04$\\
    \hline
    \multicolumn{5}{c}{At $l_{\rm cl}<1$ kpc}\\
    \hline
    All & $0.29\pm 0.04$ & $3.3_{-1.1}^{+1.8}$ & $0.19\pm0.03$\\    
    Star-forming & $0.33_{-0.07}^{+0.09}$ & $3.5_{-1.6}^{+3.7}$ & $0.23_{-0.04}^{+0.06}$\\
    Passive & $0.27_{-0.05}^{+0.06}$ & $3.5_{-1.4}^{+2.9}$ & $0.17\pm0.04$\\
    \hline
    \end{tabular}
\end{table}

The best-fit slope of $\gamma\approx 0.3$ is consistent with expectations for subsonic turbulence from the Kolmogorov theory. The estimated energy transfer rate per unit mass is $\epsilon_0\approx 0.003\,{\rm cm}^2\,{\rm s}^{-3}$ at 1 kpc, irrespective of the nature of the host galaxies (Table \ref{tab:my_label}).  Our study has yielded the first empirical constraint for the turbulent velocity field of the cool CGM around typical galaxies.  In this section, we discuss potential selection biases of the adopted sample and the implications of our findings.

\subsection{Possible sample selection bias} \label{sec:bias}

Before addressing whether or not the correlations found in our sample are due to 
sample selection biases or survey incompleteness, we first note that components detected only in \ion{H}{1} but not in any metal ions are not included in the analysis presented in this paper.  The exclusion of \ion{H}{1}-only components (which are likely extremely metal-poor or significantly more highly ionized; see below) is dictated by the need for ionic absorption lines to constrain the ionization models and to decompose thermal and non-thermal contributions.

Next, we address the $n_{\rm H}$-$l_{\rm cl}$ correlation displayed in Figure \ref{fig:nsize}.  Note that while survey sensitivity limits may prevent detection of small, low-density clumps, large, high-density clumps would have been detected if they were present.  On the other hand, Figure \ref{fig:nsize} shows an $N_{\rm H}$ threshold at $N_{\rm H}\!\approx\!10^{17}\,\cmjj$, below which no clumps are found in our sample. We note that such an $N_{\rm H}$ threshold is also seen in other CGM studies based on absorption spectra of varying qualities and ion coverage \citep[e.g.,][]{Keeney:2017,Lehner:2022}.  It is, therefore, unlikely due to the selection bias of our sample.  

To understand the origin of such a threshold, we consider the physical conditions required for 
clump formation and survival.  In a multi-phase medium, any density fluctuations would trigger condensation through the rapid cooling of high-density regions \citep[e.g.,][]{Mo:1996,Maller:2004}.  
A minimum requirement for clump formation is that the cooling time $t_{\rm cool}\sim (3/2)\, k_BT/(n_{\rm H}\Lambda)$ be comparable to or less than the sound crossing time of individual absorbing clumps, $t_{\rm cross}\approx l_{\rm cl}/c_s$, where $\Lambda$ is the cooling coefficient and $c_s$ is the sound speed of the cool gas.   When $t_{\rm cool}>t_{\rm cross}$, any density fluctuations are likely washed out by sound waves.  For clumps to form and survive, $t_{\rm cool}<t_{\rm cross}$ is preferred, leading to a minimum $N_{\rm H}^{\rm min}=n_{\rm H}\,l_{\rm cl}=(3/2)\,c_s\,(k_BT/\Lambda)$. For cool gas of $T\approx 2\times 10^4$ K (Figure \ref{fig:summary}{\it b}), $c_s\approx 20$ \kms\ and $\Lambda\sim 10^{-22}\,{\rm erg}\,{\rm cm}^3\,{\rm s}^{-1}$.  We estimate a minumun $N_{\rm H}^{\rm min}
\approx 9\times10^{16}\,\cmjj$ \citep[see also][]{Liang:2020,Faucher:2023}.  Therefore,
the absence of components below $N_{\rm H}=10^{17}\,\cmjj$ in Figure \ref{fig:nsize} can be understood as a result of this physical limit, rather than observational limitations.

In addition, we investigate whether the strong correlation between $b_{\rm NT}$ and $l_{\rm cl}$ in Figure \ref{fig:vsf} arises as a result of selection bias.  Similar to Figure \ref{fig:nsize}, we note that large clumps with small velocity widths would have been detected if they were present.  On the other hand, small clumps with broad line widths may have been missed due to S/N constraints.  Specifically, a broader absorption line covering more spectral elements would more likely be missed at a fixed signal-to-noise (S/N) ratio.  We generate synthetic absorption components for different ionic transitions, including \ion{C}{2}, \ion{O}{2}, and \ion{Mg}{2}, based on a range of gas density, metallicity, and $b$ value with a minimum total gas column per component of $N_{\rm H}=10^{17}\,\cmjj$.  For a fixed $N_{\rm H}$, we calculate the corresponding $n_{\rm H}$ for a given $l_{\rm cl}$ and compute the anticipated ionic column densities under the assumption of photo-ionization equilibrium for a range or metallicity.  Next, we generate a synthetic absorption line based on an assumed $b$ value ranging from $b<10$ \kms\ to $b\approx 100$ \kms.    We find that for gas of metallicity $0.3\,Z_\odot$, typical of the observed values in the CGM \citep[e.g.][]{Keeney:2017,Zahedy:2021,Cooper:2021}, small clumps with $l_{\rm cl}\approx 10$ pc (density $n_{\rm H}\gtrsim 0.003\,\cmjjj$ from Figure \ref{fig:nsize}) and $b\lesssim 100$ \kms\ would be detected in the survey sample.  However, reducing the metallicity to 
 $0.03\,Z_\odot$, only high-column density clumps with $N_{\rm H}>10^{18}\,\cmjj$ would be detected.

This exercise shows that our resolved CGM absorbing component sample is not biased against broad components with $N_{\rm H}$ exceeding $10^{18}\,\cmjj$ or components with $N_{\rm H}\approx 17\,\cmjj$ and a modest chemical enrichment level of $Z\gtrsim 0.1\,Z_\odot$.  While we cannot rule out the possibility of missing broad, low-column density, and metal-poor clumps, random absorber surveys have shown that broad components with $b\!>\!50$ \kms\ are rare \citep[e.g.][]{Danforth:2016,Rudie:2019,Lehner:2022}.  Therefore, we argue that the strong correlation between $b_{\rm NT}$ and $l_{\rm cl}$ is not driven by survey incompleteness.

\subsection{Implications for the thermodynamic states of the CGM}

The turbulent velocity-size relation established based on 
individually resolved cool clumps in the CGM exhibits a slope consisent with Kolmogorov theory for an isotropic, homogeneous, and incompressible fluid 
with a constant energy cascade rate. 
To drive the turbulence, a nominal source is stellar/AGN feedback, which is 
a critical ingredient in theoretical models to produce realistic-looking galaxies, transport metal from star-forming regions to low-density inter- and circum-galactic environments, and quench star formation \citep[e.g.,][]{Naab:2017}.  These feedback processes are expected to stir up the surrounding medium, disrupt continuous flows, and consequently heat up the gas \citep[e.g.,][]{Angles:2017}.  
It is, therefore, particularly interesting to see that no distinction can be found in the $b_{\rm NT}$-$l_{\rm cl}$ scaling relation between quiescent and star-forming halos at $l_{\rm cl}\!<\!1$ kpc, but the scatter is large and a notable difference between these halos is seen at $l_{\rm cl}\!>\!1$ kpc.  In particular, while the agreement with Kolmogorov theory implies that the turbulence is subsonic, \cite{Qu:2022} found that roughly $1/3$ ($\lesssim\,5$\%) of cool clumps in quiescent (star-forming) halos fall in the supersonic regime.  These super-sonic clumps, in principle, would contribute to the observed large scatter displayed in Figure \ref{fig:vsf}. It is, therefore, necessary to increase the sample size to investigate the nature of the thermodynamic state of the gas.

At $l_{\rm cl}>1$ kpc, Figure \ref{fig:vsf} shows that the $b_{\rm NT}$ vs.\ $l_{\rm cl}$ distribution for star-forming halos appears to flatten, which is also confirmed by the likelihood analysis with a substantially shallower best-fit slope when including data points at $l_{\rm cl}>1$ kpc for star-forming halos (see Table \ref{tab:my_label}).  The spatial scale at which the slope becomes flattened is often interpreted as the energy injection scale \citep[e.g.,][]{Federrath:2021}.   However, caveats remain.  In particular, for absorbers larger than 1 kpc, bulk flows within the low-density gas that could generate sufficient velocity shear become non-negligible.  In addition, using 21~cm high-velocity clouds as a guide, clouds larger than 1 kpc are uncommon \citep[e.g.][]{Hsu:2011}.  
Finally, a detailed investigation of the galaxy environment around these absorbers will help address whether or not the cloud size is overestimated due to 
the presence of local ionizing sources \citep[e.g.,][]{Qu:2023}.  

\begin{figure}
  \begin{center}
\includegraphics[width=0.475\textwidth]{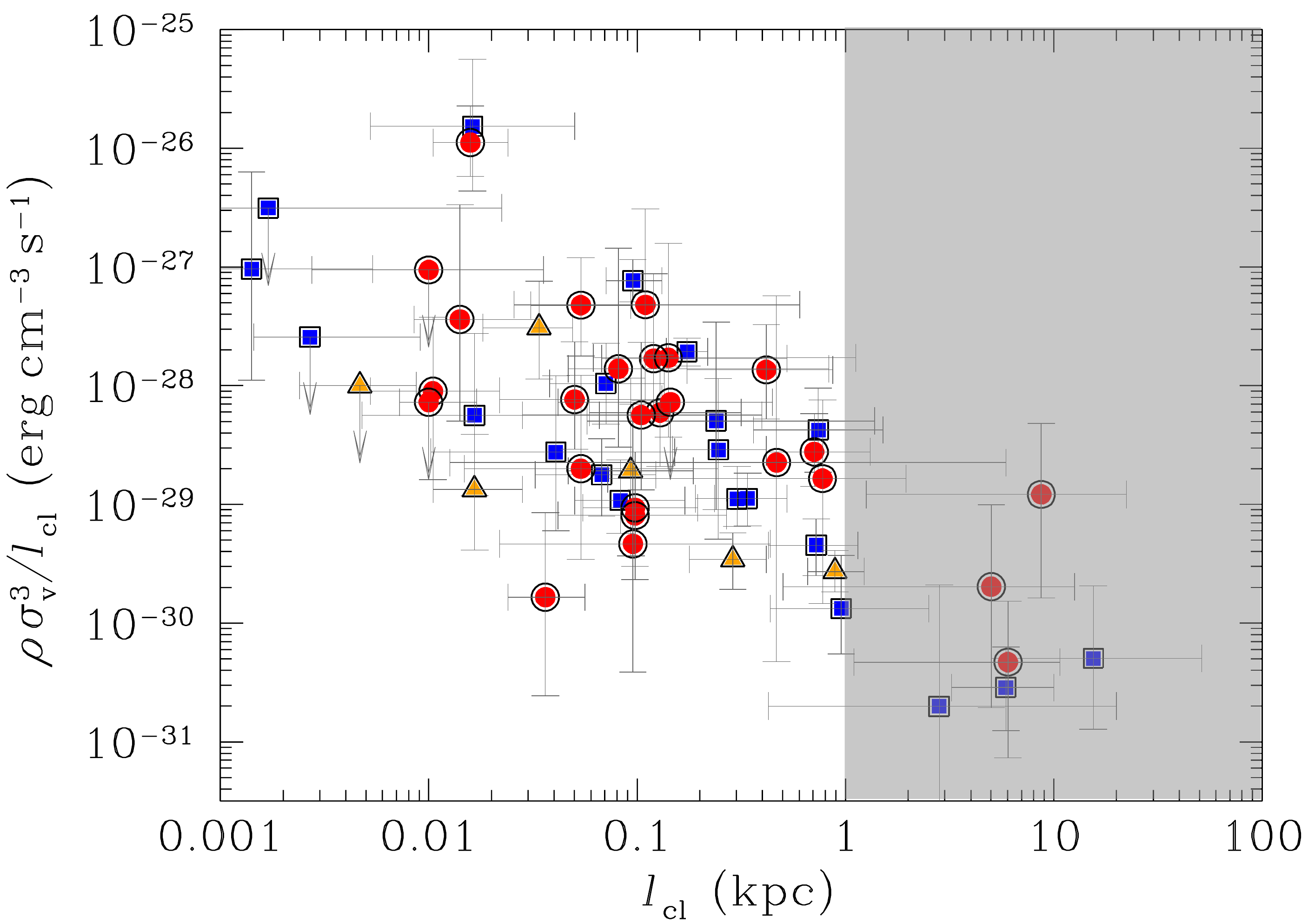}
\end{center}
\caption{Kinetic energy transfer rate per unit volume, $\rho\,\sigma_v^3/l_{\rm cl}$ versus clump size $l_{\rm cl}$.  The turbulent velocity $\sigma_v=\sqrt{3}\,\sigma_v^{\rm los}$ is computed using $b_{\rm NT}\!=\!\sqrt{2}\,\sigma_v^{\rm los}$. Different symbols indicate the sources of measurements described in Figure \ref{fig:summary} and in the main text.  Following Figure \ref{fig:summary}, upper limits (downward arrows) arise as a result of non-constraining $b_{\rm NT}$.  The parameter space at $l_{\rm cl}>1$ kpc is greyed out due to the uncertain nature of $b_{\rm NT}$.  Over the range of spatial scales from $l_{\rm cl}\approx 1$ pc to $\approx 1$ kpc, the energy transfer rate per unit volume displays a mildly increasing trend toward smaller scales with a median value of $\approx 6\,\times\,10^{-29}\,{\rm erg}\,{\rm cm}^{-3}\,{\rm s}^{-1}$ at $l_{\rm cl}\approx 100$ pc and an order of magnitude scatter.  
}
\label{fig:ke}
\end{figure}

At $l_{\rm cl}\lesssim 1$ kpc, we infer a roughly constant energy transfer rate per unit mass of $\epsilon\approx 0.003\,{\rm cm}^{2}\,{\rm s}^{-3}$, which is comparable to what is seen in high-redshift \ion{C}{4} absorbers \cite{Rauch:2001}. Given the best-fit densities of individual clumps from the photo-ionization models, we also evaluate the total kinetic energy transfer rate per unit volume for these clumps, $\rho\,\sigma_v^3/l_{\rm cl}$, over the range of spatial scales from $l_{\rm cl}\approx 1$ pc to $\approx 1$ kpc. These are displayed in Figure \ref{fig:ke}.  Despite an apparent inverse correlation between the energy transfer rate per unit volume and clump size in Figure \ref{fig:ke}, we note that this is largely driven by the points at $l_{\rm cl}>1$ kpc (shaded area) for which the nature of $b_{\rm NT}$ remains ambiguous. A generalized Kendall's test using all the points at $l_{\rm cl}<100$ kpc, including upper limits, returns a rank-order coefficient $\tau=-0.36$ and a $p$ value of 99.9\%.  Excluding the points at $l_{\rm cl}>1$ kpc reduces the correlation coefficient to $\tau=-0.26$.  

For clumps of typical size $l_{\rm cl}\!\approx\!100$ pc, the energy transfer rate per unit volume is low with a median value of $\sim\!6\!\times\!10^{-29}\,{\rm erg}\,{\rm cm}^{-3}\,{\rm s}^{-1}$. In contrast, in the neutral interstellar medium, the energy transfer rate per unit mass at 100 pc is $\epsilon\!\approx\!0.02$ - $0.1\,{\rm cm}^2\,{\rm s}^{-3}$ 
with a roughly constant energy transfer rate per unit volume of $\sim\!10^{-25}\,{\rm erg}\,{\rm cm}^{-3}\,{\rm s}^{-1}$ over a broad range of scales from 0.01 pc to 100 pc \citep[ISM, e.g.,][]{Hennebelle:2012,Arthur:2016}.  Despite a large scatter, the turbulence observed in the cool CGM, on average, appears to be significantly more quiescent than in star-forming regions.

At the same time, the time scale over which the kinetic energy is dissipated is relatively short, $\lesssim 100$ Myr.  With a low level of turbulence and a relatively short dissipation time scale, it is unlikely that the gas is stirred up by recent energetic outflows but nearby sources are still needed to continue to inject energy into the gas.  One possibility is fragmentations of accreted gaseous streams from the intergalactic medium \citep[e.g.,][]{Vossberg:2019,Aung:2019}.  Alternatively, merging satellites and satellite interactions in galaxy halos, both around quiescent and star-forming galaxies, are promising candidates for maintaining such energy input in both star-forming and quiescent halos \citep[e.g.,][]{Tal:2012}.  Indeed, many of these cool CGM clumps are found in a group environment \citep[e.g.,][]{Cooper:2021,Qu:2023}.  

We note two outliers at $l_{\rm cl}\approx 10$ pc in Figure \ref{fig:ke}, which have an inferred energy transfer rate per unit volume exceeding $\approx\!10^{-26}\,{\rm erg}\,{\rm cm}^{-3}\,{\rm s}^{-1}$.  These clumps also have among the highest $n_{\rm H}$ of the sample 
(see Figure \ref{fig:nsize}).  The high energy transfer rate per unit volume implies a much faster time scale for energy dissipation, suggesting possibly being stirred up by different processes.  


Recently, \cite{Chen:2023} analyzed IFS observations of spatially-extended line-emitting nebulae around four QSOs at $z\approx 0.5$-1.  While three of the four cases did not have a sufficiently large dynamic range in spatial scales to provide meaningful constraints on the VSFs, the largest QSO nebula in their sample exhibits a VSF that was in remarkable agreement with the Kolmogorov energy cascade from $\approx 40$ kpc to $<10$ kpc at an extremely high energy transfer rate per unit mass of $\epsilon\approx 0.2\,{\rm cm}^{2}\,{\rm s}^{-3}$.  The implied subsonic turbulence in the vicinity of a luminous QSO is puzzling.  High-resolution absorption spectroscopy to resolve individual clumps in these QSO nebulae and to constrain the internal gas dynamics is needed to gain physical insight into the origins of the turbulence observed in QSO host halos. Such observations will also facilitate a direct comparison with studies in halos around typical galaxies.

\begin{acknowledgments}
    
We thank the anonymous referee for the swift review and constructive comments that helped improve the presentation of the paper. HWC thanks Fausto Cattaneo and Andrey Kravtsov for many discussions that helped shape the analysis presented here.  HWC, ZQ, and MC acknowledge partial support from NSF AST-1715692, HST-GO-15163.01A, and NASA ADAP 80NSSC23K0479 grants.  EB acknowledges support by NASA under award number 80GSFC21M0002.  SC gratefully acknowledges support from the European Research Council (ERC) under the European Union's Horizon 2020 Research and Innovation programme grant agreement No 864361.  CAFG was supported by NSF through grants AST-2108230 and CAREER award AST-1652522; by NASA through grants 17-ATP17-0067 and 21-ATP21-0036; by STScI through grant HST-GO-16730.016-A; and by CXO through grant TM2-23005X.  SL acknowledges support by FONDECYT grant 1231187.

\end{acknowledgments}

\bibliography{main} 
\bibliographystyle{aasjournal}



\end{document}